\documentstyle{BSAXwork}

\input epsf.sty

\def \AAP #1 #2 {{\em Astron. Astrophys.\/} {\bf #1}, #2}
\def \AAL #1 #2 {{\em Astron. Astrophys. Lett.\/} {\bf #1}, L#2}
\def \AAR #1 #2 {{\em Astron. Astrophys. Rev.\/} {\bf #1}, #2}
\def \AAS #1 #2 {{\em Astron. Astrophys. Suppl. Ser.\/} {\bf #1}, #2}
\def \AJ #1 #2 {{\em Astron. J.\/} {\bf #1}, #2}
\def \ANNREV #1 #2 {{\em Ann. Rev. Astron. Astrophys.\/} {\bf #1}, #2}
\def \APJ #1 #2 {{\em Astrophys. J.\/} {\bf #1}, #2}
\def \APJL #1 #2 {{\em Astrophys. J. Lett.\/} {\bf #1}, L#2}
\def \APJS #1 #2 {{\em Astrophys. J. Suppl.\/} {\bf #1}, #2}
\def \APSS #1 #2 {{\em Astrophys. Space Sci.\/} {\bf #1}, #2}
\def \ASR #1 #2 {{\em Adv. Space Res.\/} {\bf #1}, #2}
\def \BAIC #1 #2 {{\em Bull. Astron. Inst. Czechosl.\/} {\bf #1}, #2}
\def \JSQRT #1 #2 {{\em J. Quant. Spectrosc. Radiat. Transfer\/} {\bf #1}, #2}
\def \MN #1 #2 {{\em Mon. Not. R. Astr. Soc.\/} {\bf #1}, #2}
\def \MEM #1 #2 {{\em Mem. R. Astr. Soc.\/} {\bf #1}, #2}
\def \PLR #1 #2 {{\em Phys. Lett. Rev.\/} {\bf #1}, #2}
\def \PASJ #1 #2 {{\em Publ. Astron. Soc. Japan\/} {\bf #1}, #2}
\def \PASP #1 #2 {{\em Publ. Astr. Soc. Pacific\/} {\bf #1}, #2}
\def \NAT #1 #2 {{\em Nature\/} {\bf #1}, #2}
\def \SAIT #1 #2 {{\em Mem.\ Soc.\ Astron.\ It.\/} {\bf #1}, #2}
\def \MESS #1 #2 {{\em The Messenger\/} {\bf #1}, #2}
\def \ASTRNACH #1 #2 {{\em Astron. Nach.\/} {\bf #1}, #2}
\def \RVMPH #1 #2 {{\em Rev. Mod. Phys.\/} {\bf #1}, #2}
\def \APH #1 #2 {{\em Astropart. Phys.\/} {\bf #1}, #2}

\begin{opening}

\title{A SSC model for the spectral variability of the intermediate blazar ON 231}
\author{S. Ciprini$^{1,2}$}
\institute{$^1$Physics Department, University of Perugia, Italy\\
$^2$Astronomical Observatory, University of Perugia, Italy}
\date{} 
\end{opening}

\begin{document}

\oddpagefooter{}{}{} 
\evenpagefooter{}{}{} 
\medskip  

\begin{abstract} 
In April-May 1998 the blazar ON 231 (W Com, B2 1219+28) showed an
extraordinary optical outburst alerted by the Perugia Astronomical
Observatory. The source reached the most luminous state, observed
at least since the beginning of the previous century.
\textit{Beppo}SAX observed ON 231 in the band from 0.1 up to 100
keV, detecting an X-ray spectrum with a peculiar concave profile,
that could be considered as a signature of the intermediate blazar
subclass. The spectral energy distributions (SED) of blazars were
studied with a time dependent synchrotron self-Compton (SSC)
model, tested here on this object. Even with the preliminary
``toy'' model of a single flaring region, emitting by means of
mere SSC process, the X-ray spectrum is reasonably reproduced. The
SED was calculated with a hard spectrum of the injected electrons
and with other physical parameters comparable to previous works,
validating a base consistence of the method.
\end{abstract}

\medskip
\section{Introduction}
The overall spectral energy distribution (SED) of blazars show a
two-bump structure where the lower frequency hump is peaked either
in the IR/optical (low frequency peaked LBL or ``red'' blazar) or
in the UV/X-ray bands (high frequency peaked HBL or ``blue''
blazars), (Padovani \& Giommi 1995, Ghisellini et al. 1998, Urry
1999) and is believed to be produced by synchrotron emission,
while the higher frequency component should be given by inverse
Compton (IC) scattering. Synchrotron radiation and IC scattering
was usually interpreted as due to diffusive shock acceleration of
charged particles within a plasma jet, which itself moves at
relativistic speed and point toward the observer. ON 231 is one of
the best examples in which both the synchrotron decline tail and
the further rise of the IC bump in the broadband spectrum, were
observed simultaneously at the X-ray wavelengths. The consequent
concave shape of the X-ray spectrum detected by \textit{Beppo}SAX
(Tagliaferri et al. 2000) could be the important signature of the
intermediate blazar subclass (i.e. the transition from LBL to
HBL). Detailed studies on intermediate objects are fundamental to
probe the continuous spectral sequence of blazars ``flavours''
resulting from the source power dependence and from the
distribution of the intrinsic physical parameters (Fossati et al.
1998).
\section{The outburst of April-May 1998}
The blazar ON 231 (W Com, B2 1219+28; z=0.102) was one of the
first radio sources classified as a classical BL Lacertae object
(Biraud 1971; Browne 1971). Perugia Observatory, equipped with a
0.4 m robotic telescope (Tosti et al. 1996), since 1994 is
carrying out an intensive optical monitoring of ON 231. The
optical flux was at an higher level respect to historical mean
value, showing in each season a superimposition of many flares
with different amplitudes and time ranges (Tosti et al. 1998). In
April 1998 the source had an extraordinary outburst (Massaro et
al. 1999) and reached the highest brightness level ever recorded
since the Wolf's discovery of the optical counterpart (Wolf 1916).
In the maximum phase the magnitude of ON 231 was R(Cousins)=12.2
(density flux $F_{\nu}= 42.9$ mJy, or $\log_{10}(\nu F_{\nu})=
-9.69$ erg s$^{-1}$ cm$^{-2}$, corrected for Galactic extinction).
The linear optical polarization also increased suddenly, giving
evidence of the pure non-thermal origin of the flare. 1999 VLBI
maps displayed a new prominent two-sided structure, suggesting a
relation between the optical enhanced activity and changes in the
innermost radio structure (Massaro et al. 2001). The subsequent
optical behaviour showed a slow decline in the mean luminosity
(Tosti et al. 2002). \textit{Beppo}SAX observed ON 231 in May 1998
with a good sensitivity and spectral resolution, following the
extraordinary optical outburst and detecting the unexpected hollow
X-ray spectrum between the two bumps of the SED. This data was
used to test and probe the parameter plausibility and the
robustness of a mere synchrotron self-Compton (SSC) time-dependent
model. In fact this spectrum shape is difficult to fit for a
preliminary ``one--zone--one--process'' implementation, without
any external IC contributes.
\begin{figure}[t]
\begin{center}
\leavevmode
\epsfysize=6cm 
\hspace{-5.mm}\vspace{-0.2cm} \epsfbox{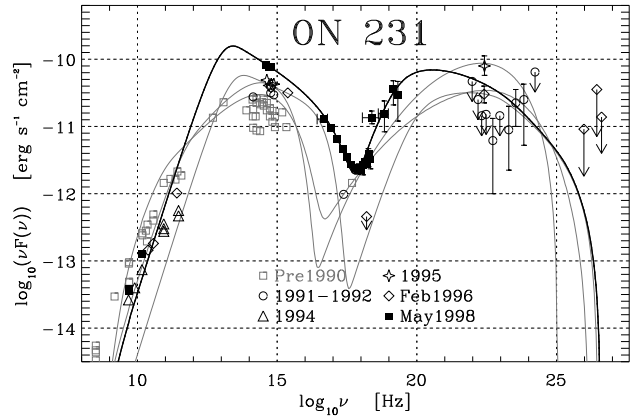}
\caption[h]{Attempts to fit the SED of ON 231 (z=0.102) with a
single flaring plasma knot using a pure SSC model. The darkest fit
in the picture correspond to the concave X-ray spectrum detected
in May 11-12, 1998 by \textit{Beppo}SAX (data was taken from
Tagliaferri et al. 2000) after an exceptional optical outburst
occurred in April. We report for this SED the following
parameters: injected electron Lorentz factors $\gamma_{min}=500$,
$\gamma_{max}=10^{5}$, electron number density $N=10^{6}$
cm$^{-3}$, tangled magnetic field intensity $B=0.6$ G, injected
electron power law index $s=-3.4$, flaring region dimension
$R=4\times10^{16}$ cm, electron escape time $t_{esc}=1.5~R/c$,
Doppler beaming factor ${\mathcal D}=10.5$. All optical post-1994
data are taken from the Perugia Observatory
database.}\vspace{-0.2cm}\label{fig:fiton231}
\end{center}
\end{figure}
\section{Model and simulations}
An electron plasma confined by a magnetic field into a non-thermal
flaring knot in blazar jets, can be described, in accord to a
statistical reduced description, with a density operator or a
one--particle distribution function giving the particle number
density in the phase space. An appropriate truncation of the exact
kinetic relations and a diffusion approximation for system of
particles moving under stochastic forces that keep producing small
changes in particles dynamics can used to further simplify the
kinetic equations (Chandrasekhar 1943). Taking an isotropic and
space homogeneous distribution function and integrating in volume
and angular momentum-space coordinates, the dependency from pitch
angle, spatial and spherical momentum--space coordinates can be
removed. Thus may be used a simple one--dimensional and modified
Fokker--Planck equation for the number particle density
$N(t,\gamma)$ [cm$^{-3}$] (Lorentz $\gamma=E/m_{e}c^{2}$ is the
dimensionless factor) in order to describe the time--dependent
evolution of the electron distribution in the flaring volume. This
equation can be though as a random walk in the energy space of
electrons and can be written as
\begin{equation}\label{fokkerprinc}
      \frac{\partial N(t,\gamma)}{\partial t}=\frac{1}{s(\gamma)}\frac{\partial}
      {\partial \gamma}\left(D(t,\gamma)\frac{\partial N(t,\gamma)}
      {\partial \gamma } + A(t,\gamma) N(t,\gamma) \right)
      - \frac{N(t,\gamma)}{\tau(\gamma)}+Q(t,\gamma),
\end{equation}
where $D(t,\gamma)$ represent the sum of the various energy
diffusion coefficients; $A(t,\gamma)$ is the advective term
contains energy gains and losses; $\tau^{-1}(\gamma)$ is the
probability of particle disappearance by escape, $Q(t,\gamma)$
[cm$^{-3}$ s$^{-1}$] is the source term, while $s(\gamma)$ is a
scaling factor for the volume element. The solutions of this type
of equations are very sensitive to initial and boundary conditions
and to injection and losses forms.\par In this preliminary SSC
model, a single homogeneous one--particle plasma region was
considered (as done in Chiaberge \& Ghisellini 1999), subordinated
to injection of accelerated relativistic electrons with a
power-law energy distribution and to pure radiation cooling. An
uniform tangled magnetic field $\textit{\textbf{B}}$ was
considered, even if a superposed small-amplitude plasma turbulence
(B\"{o}ttcher et al. 1997), needs to be implemented for more
realistic simulations. The ensemble synchrotron spectrum was
integrated by the calculated electron distribution at any given
time. The IC spectrum result as the interaction of the energetic
electron distribution with the synchrotron photons using a
standard technique (Band \& Grindlay 1985, Li \& Kusunose 2000).
The produced spectra were transformed to the observer using the
Doppler beaming factor ${\mathcal D}=((1+z)\Gamma(1-\beta \cos
\theta))^{-1}$ (where $\Gamma$ is the bulk Lorentz factor of the
emitting region, $\theta$ the angle respect to the observer) and
using the luminosity distance of the source.
\begin{figure}[t]
\begin{center}
\leavevmode
\epsfysize=4.cm 
\hspace{-1.mm} \vspace{-0.2cm}\epsfbox{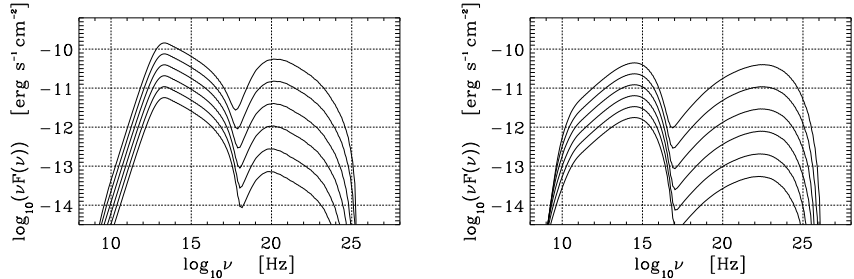}
\caption[h]{An example of pure cooling temporal evolution in units
of the light crossing time. Left: for the May 1998 darkest black
fit of Figure 1. Right: for the 1991-1992 data of Figure
3.}\vspace{-0.2cm}
\end{center}\label{fig:on231tempevoluz}
\end{figure}
\par The concave X-ray spectrum of ON 231
measured from 0.1 keV up to 100 keV by \textit{Beppo}SAX in May
11-12, 1998, and the simultaneous optical multiband data of
Perugia Observatory (May 11 $R_{c}= 13.06$, May 12 $R_{c}=13.25$)
was fitted with the physical parameters reported in the caption of
Figure 1, that are in rough agreement with the value calculated in
the paper of Tagliaferri et al. (2000). Only the flaring region
dimension $R$ was taken almost one order of magnitude greater than
the value reported in the paper. A single no-break power-law for
the injected electron energy distribution is able to reproduce the
X-ray spectrum, with a very steep power index ($Q(\gamma)\propto
\gamma^{-3.4}$ between $\gamma_{min}$ and $\gamma_{max}$). This
reflects the common behaviour of harder injection spectrum during
flaring states. The steeper index counterbalances the larger
volume of the flaring region, giving the same global energetics.
The IC gamma--ray part of the SED is into some limits and
detections of EGRET in 1991--1992 and February 1996, even if this
data are indeed weak constraints for the May 1998 behaviour. A
weakest point of the simulation, could be the predicted relevant
synchrotron peak at mid--IR wavelength, because it assumes a flux
much higher than the unique available far--IR data (IRAS) about ON
231. On the other hand this could be the result of the April
exceptional outburst that raised the non-thermal IR-optical-UV
synchrotron continuum to the highest level recorded since the
beginning of the previous century ($F_{\nu}>50$ mJy in $I_{c}$
band in April 23, 1998). This is witnesses by the flattest optical
spectral distribution ever observed during the maximum (Massaro et
al. 1999), that may be extended to IR wavelengths. The
subsequently faster cooling of the optical emission respect to the
mid--IR photons, should have dropped the optical flux until the
photometric values registered almost 20 days later, contemporary
with the SAX detections, while a larger amount of infrared flux
could have survived for more time.
\begin{figure}[t]
\epsfysize=4.cm 
\hspace{-3.mm} \vspace{-0.2cm}\epsfbox{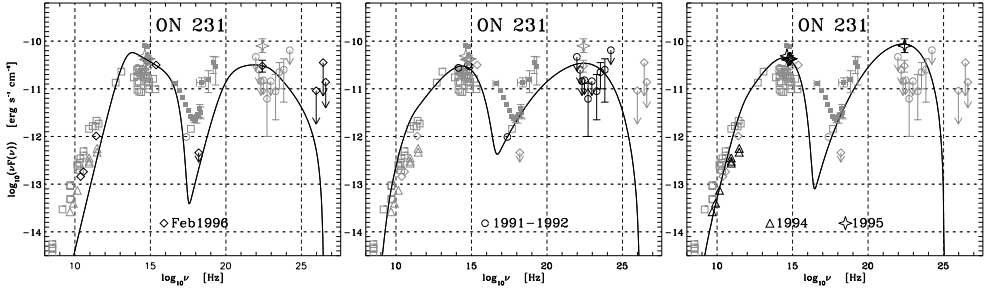}
\caption[h]{Other SEDs of ON 231 produced with the model for
simultaneous and not-simultaneous data before
1998.}\vspace{-0.2cm}\label{fig:other3fiton231}
\end{figure}
At longer wavelengths the ``toy'' model of a single flaring blob,
with this hard injection, do not reproduce the spectrum, probably
due also to stationary radio--emission in the the underlying jet.
In Figure 2 is shown an example of plain cooling, plus escape,
temporal evolution for the May 1998 SED and the 1991-1992 spectrum
(see Figure 3), calculated by the time--dependent simulation. In
Figure 3 are displayed three spectra for simultaneous data of
February 1996 and for other not contemporary observations. Here
the profiles of the SED are more similar to the HBL spectrum
shapes, generally simpler to simulate with a plain SSC model. All
this other SEDs passes through the IRAS data and were produced
with plausible parameters. The model seems physically acceptable
also in this cases, even if this data have a little means,
compared to the strict constraints given by the excellent
observations of \textit{Beppo}SAX.





\end{document}